\documentclass[twocolumn,showpacs, prb,aps]{revtex4}

\begin{document}
\date{\today}
\title{Criticality in strongly correlated fluids}
\author{\bf Yan Levin} 
\affiliation{\it Instituto de F\'{\i}sica, Universidade Federal
do Rio Grande do Sul\\ Caixa Postal 15051, CEP 91501-970, 
Porto Alegre, RS, Brazil\\ 
{\small levin@if.ufrgs.br}}

\begin{abstract}

In this brief review I will discuss criticality
in strongly correlated fluids.  
Unlike simple fluids, molecules of which
interact through short ranged isotropic potential,
particles of strongly correlated fluids usually interact
through long ranged forces of Coulomb or dipolar form.
While for simple fluids mechanism of phase separation
into liquid and gas was elucidated by van der Waals more
than a century ago, the universality class of
strongly correlated fluids, or in some cases even 
existence of liquid-gas phase separation remains uncertain.

\end{abstract}
\maketitle
\bigskip


Over a century ago van der Waals has provided a basic
explanation of what drives the liquid-gas phase separation
in simple molecular fluids~\cite{Waa73}. 
This work had its lasting value
not because of its quantitative accuracy, which has since
been  surpassed by more accurate empirical equations
of state, but
because of a  simple physical picture of criticality that
it provided. The fundamental
insight of van der Waals was to 
modify the ideal gas equation of state by 
including two fundamental effects:

\begin{enumerate}
\item The hardcore for each molecule.
\item The short ranged intermolecular  interaction. 
\end{enumerate}

Thus, van der Waals suggested to modify the ideal gas equation
of state by replacing the total volume $V$ in the $PV=nRT$ 
by an {\it effective} volume available to a 
given molecule, $V_{\rm eff}=V-bN$, where $b$ is a factor
proportional to molecular
volume and $N$ is the total number of particles. This
means  that because of a hardcore repulsion, each particle moves
not inside the total volume $V$ but in the reduced volume 
$V_{\rm eff}$, which excludes the space 
occupied by other particles.  
Furthermore, he realized that in order to keep cohesiveness of
the fluid, molecules  must attract one another. Not, knowing
the precise nature of intermolecular forces, but based on a
physical insight and understanding of thermodynamics,
van der Waals argued that the forces must be  
short ranged. Actually the idea of hard impenetrable atoms
interacting by short ranged attractive forces can be traced
all the way to Newton.  However, in his writings Newton was 
never fully committed to this position. Depending 
of the subject he switched between the atoms which
interacted with short ranged attractive forces~\cite{New52} 
to atoms which repel
their nearest neighbors~\cite{New87} with a force proportional
to $1/r$.  The first clear commitment to short ranged 
attractive interatomic forces appeared in the 
work of Laplace~\cite{Lap22}
in 1822.
He used the existence of these forces 
to account for the surface tension and
capillarity of liquids. 
Laplace believed that attractive
forces acted when the atoms were close together --- as
in solid or liquid --- while the 
repulsive forces dominated
in gases.  Notion that the nature of atoms and the
forces with which they interact 
depends on the thermodynamic 
phase, i.e. solid, liquid, or gas,  
seems to have dominated much of the early 
19th century atomic 
theory~\cite{Bru99}. 
The idea that all three states of matter should be
explained with the same atomic model, did not take hold until
the work of van der Waals in 1873~\cite{Bru99}.

Of course, if the attraction between particles exists,
it must reduce the
pressure $P$ bellow that of an ideal gas at the same
density, so that $P=P_{id}-\Delta P$. Based on 
kinetic theory introduced earlier by 
Clausius~\cite{Cla57} and Maxwell~\cite{Max67},
Van der Waals 
suggested that the ideal gas equation of state
still applies, but in the form $P_{\rm id}V_{\rm eff}=nRT$. 
The question, however remained: what is  
$\Delta P$?

To answer this question van der Waals proposed what is now
called the `` mean-field field approximation''.  
Since forces
between particles are short ranged, 
each molecule interacts only
with its neighbors, i.e. with the 
particles which
are within some radius $\delta$ from it. Now, how many molecules
are within distance $\delta$ of any given particle?  
Since on average  
particles are uniformly distributed with 
density $\rho=N/V$, the number of molecules
within radius $\delta$ of a given 
particle is $\rho 4 \pi \delta^3/3$.
It is reasonable to think of $\Delta P$ as an effective 
energy density of interaction.
Therefore, we conclude that
\begin{equation}
\label{1}
\Delta P=\frac{\epsilon N \rho}{2 V} \frac{4 \pi \delta^3}{3}\;.
\end{equation}
In this formula $\epsilon$ is the energy of  
''pair interaction'' and the factor $1/2$ is to account
for the double counting.  Substituting this into
the modified ideal gas equation we are led
to the celebrated van der Waals equation of state $(vdW)$,
\begin{equation}
\label{2}
P+\frac{a }{v^2}=\frac{k_BT}{(v-b)}\;,
\end{equation}
where $v$ is the volume per particle, 
$a=2 \pi  \epsilon \delta^3/3$, and $k_B$ is
the Boltzmann constant.

The derivation above also makes it clear  what
are the main effects omitted on the way to Eq.(\ref{2}).  
By concentrating only on the average distribution
of particles, we have neglected the role of correlations. 
It is evident that
if we fix one particle and look at the distribution
of other molecules around it, due to
attractive forces there will be an enhanced probability
of finding another molecule in the vicinity of the fixed particle.  
Certainly this
kind of effect is not accounted for 
in the $vdW$ equation of state. For a class
of ''simple'' fluids the correction due to 
positional correlations between
particles, however, is not very
important and does not change, 
beyond quantitative, the basic
predictions of the theory.  

The $vdW$ equation provides
us with an insight into the mechanism of
liquid-gas phase separation.  
The phase transition, 
appears as a violation of thermodynamic inequality
$\partial P/\partial \rho \ge 0$ bellow some critical
temperature $T_c$.
Also the free energy must be a convex function
of density.  If convexity is violated it leads to appearance
of  two phases --- liquid and gas --- which together
have the free energy lower than the ''anomalous'' 
homogeneous state.
It took further work of Maxwell, Boltzmann, 
and Gibbs~\cite{Gib28} to clarify
the details of this seemingly simple observation.  
The basic picture of fluid criticality, however, remained
unchallenged for seventy years. 
This situation changed 
when the work of Onsager showed that 
there was a significant flaw in the  
mean-field picture based on the $vdW$ equation.  Onsager
was able to solve exactly a related model of appearance of 
magnetization in ferromagnetic materials~\cite{Ons44}.  What he found
was that near the critical point, magnetization behaved very
differently from the prediction of a mean-field theory.  The
situation remained unclear until 
the pioneering work of Wilson,
Fisher and the others, 
in the early 1970's, almost exactly hundred
years after the original work of van der 
Waals~\cite{Wil71,Wil72}.  These authors
showed the essential role played by thermal fluctuations
in the vicinity of a critical point, and provided an
explanation for the  existence of various 
universality classes
of critical behavior.  These universality classes are 
characterized by the numerical value of  
critical exponents associated with the second order 
phase transitions. It was found that 
very different systems can belong
to the same universality class. 
In particular, it was demonstrated 
that the liquid-gas criticality
belongs to the same universality class as the 
paramagnetic-ferromagnetic transition in metals. This universality
class is now called Ising. 
Furthermore, all
systems in which interactions between the
particles are short ranged, and the
Hamiltonian does not posses specific symmetries, 
belong to the same Ising universality class.
Recently, however,  something very surprising
has happened.  A new universality 
class of critical behavior has been observed.  
It was first noticed in the critical behavior
of solutions containing organic salts~\cite{Pit90}. 
Thus, it was found
experimentally that the criticality of systems in which
the Coulomb force plays 
the dominant role appeared to exhibit 
mean-field critical behavior.   At first
this observation might not sound very surprising, 
after all the
Coulomb force is extremely long ranged, while the Ising
universality class is specifically for systems in which the
interactions are short ranged.  On second thought
the situation does not appear
so clear.  Although the {\it bare} Coulomb force 
is extremely long ranged, the {\it effective} interaction
between any two particles of an electrolyte 
is dynamically screened by the other ions 
and is short ranged~\cite{Deb23}.  Furthermore,
all the 
theoretical arguments appear to converge to conclusion
that criticality in a Coulomb fluid
should not be any different
than in systems dominated by short ranged forces~\cite{Fis93,Lev96,Lee96}.
Specifically, it is possible to characterize the width of
a critical region, in which  scaling 
is observed, by a so called Ginsburg parameter.  
All theoretical
arguments suggest that the critical region of 
a Coulomb fluid 
should be comparable to that of a 
simple van der Waals fluid~\cite{Fis93,Lee96}.  
Experimentally, however, if one insists 
on fitting the data to Ising universality class, one finds
that the Ising 
critical region is at least two orders of magnitude narrower
than the critical region for a simple fluid with short 
ranged interactions~\cite{Fis94}.  
Outside this region the the critical behavior 
appears completely mean-field-like.
The theoretical challenge is to account for the mean-field
criticality observed in organic electrolytes, or if the
critical behavior is convincingly proven to be 
Ising, to justify the narrowness of
the Ising critical region.

There is a striking difference between the mechanism
of phase separation in a 
van der Waals fluid and that in a Coulomb fluid.
Recall that the phase separation in a $vdW$ fluid
is driven by the competition between hardcore repulsion
and interparticle attraction.  Contrary to this,
the hardcore repulsion plays only a small role in 
coulombic criticality~\cite{Fis93}. 
The reason for this is that
for electrolytes and molten salts 
the critical point is located at very low reduced
density and temperature, about a factor of ten lower 
than these
values for a $vdW$ fluid~\cite{Val91}.  
Clearly at such large separations,  hardcore
is only a marginal perturbation.  Based on the pioneering
ideas of Debye, H\"uckel, and Bjerrum~\cite{Deb23,Bje26}, 
it is possible
to construct a theory which accounts quantitatively for
the location of critical point in a Coulomb fluid~\cite{Fis93,Lev96}.
Within this theory one can explicitly see that the
hardcore, free volume term, which is so essential to the
understanding of criticality in simple fluids, is 
completely negligible for Coulomb systems.  In fact
the phase separation in these systems is purely 
the result of an
electrostatic instability~\cite{Fis93}.  If one studies 
coulombic criticality, one also notices
another striking difference from systems with isotropic
short ranged attractions.  Let us consider a simple
model of an electrolyte idealized by the restricted primitive
model $(RPM)$.  Within the $RPM$, ions are
treated as an equal-sized hard spheres of positive and
negative charge, and solvent as a uniform dielectric background. 
We shall restrict
our attention to charge neutral systems.  The condition
of charge neutrality is essential in order to have  
thermodynamic limit. 
Since on average the particles of an electrolyte are
uniformly distributed throughout the volume, 
it is evident that the average electrostatic 
potential 
is zero. This means that  
the mean-field correction to ideal gas 
pressure is also zero!  The fact
that the $RPM$ does have liquid-gas phase transition must, therefore, be
attributed to the  effects not present 
at the mean-field  level.     
These effects are the correlations between 
positive and negative
ions, neglected within the mean-field approximation. 
At very low reduced
temperatures, at which  phase separation occurs,  
$RPM$ is a strongly correlated fluid. 
It is precisely these correlations which drive the 
liquid-gas phase separation~\cite{Fis93,Lev96}.
 
Another very interesting example of a highly 
correlated fluid is a system
of dipolar hard spheres $(DHS)$~\cite{Gen70,Wer71,Rus73,Gro96,Sea96,Roi96,Tav97,Lev99,Lev01}.  This is probably the simplest 
model of a polar fluid.  Indeed, it was expected that
as the temperature is lowered, the $DHS$ should
phase separate into coexisting 
high density liquid and  low density
gas phases~\cite{Gen70,Wer71,Rus73}. 
It came, therefore, as a big surprise
when the simulations in the early $90's$ could not locate
the anticipated liquid-gas 
transition \cite{Wei93,Lee93,Ste94}.
Instead, as the temperature was lowered, 
the dipolar spheres 
aligned forming weakly interacting 
polymer-like chains.  It was argued that
formation of these chains prevented condensation 
of $DHS$~\cite{Lev99}.

In fact, it is once again
easy to see that the dominant correction to the ideal
gas behavior is due to  
correlations between  particles.  Evidently 
in a fluid phase, all the particles are 
uniformly distributed throughout the
system, with dipolar 
orientation not having any preferred direction.
This implies that the average electric field inside a dipolar
fluid is zero, and the mean-field contribution to pressure
is, once again, zero. 
 The dominant
contribution to the excess free energy is  
due to correlations between the particles and not due 
to mean-field,
as was the case for a simple van der Waals fluids.  
In fact,
we find that the correlations, in this case, 
are so strong as to wash out the 
phase separation altogether\cite{Lev99}. 
That is the dipoles align forming
weakly interacting chains, instead of phase separating into
coexisting high and low density phases. 
What is even more interesting is that
the usual liquid-gas phase stransition can be restored
by adding to the $DHS$ some dispersion interaction of the
van der Waals, $1/r^6$, form~\cite{Lee93,Lev01}. 
It can be shown that
depending on the amplitude of dispersion, as
compared to dipole-dipole interaction, 
a liquid-gas phase transition
can either be absent or present. We shall denote 
the relative strength of  dispersion force as
it compares to the dipole-dipole interaction by a
parameter $\lambda$.  Thus, when $\lambda$ is near
unity, the two interactions are more or less
comparable in strength.  When $\lambda=0$ the usual
$DHS$ model is recovered.   When $\lambda=1$ we find
that there is a phase transition, location of which
can be predicted, with about $30\%$ accuracy, 
using a 
``correlated mean-field theory''~\cite{Lev99,Lev01}.  
This theory combines into a single free energy 
contributions from both the isotropic van der Waals 
force and from the dipolar correlations. 
The critical point is  obtained from the
study of convexity of this free energy. It is important
to stress that the correlated mean-field theory still does
not take into account the thermal fluctuations important
near the critical point.

As the relative strength of the dispersion force decreases, 
and the
net interaction between  particles 
becomes dominated by the long
ranged dipole-dipole force, 
we find that the accuracy of 
the theory improves dramatically.
In fact for $ 0.3<\lambda<0.5$ the values for the critical
temperature predicted by the theory 
are in perfect agreement with the Monte Carlo
simulations~\cite{Lev01}. At $\lambda \approx 0.3$ the
phase separation disappears both in the simulations and 
in theory~\cite{Lee93,Lev01}.

The fact that a theory which does not take 
into account thermal
fluctuations can predict location of a critical point
with a high degree of accuracy, carries some very profound 
implications.  We note that the accuracy of the
theory increases as $\lambda$ decreases. As mentioned
already, for  $\lambda=1$
disagreement between the theory and the  simulations
is about $30 \%$.  This value is very much comparable
to the accuracy of a mean-field theory for fluids
with short ranged dispersion
force, such as argon.  It is well known that the thermal
fluctuations tend to depress the critical temperature
by about this amount.  What is interesting is that as 
the weight of an isotropic short ranged interaction is 
decreased --- by diminishing the value of $\lambda$ --- and 
the effective interaction becomes longer
ranged, the agreement between 
simulations and theory improves significantly~\cite{Lev01}.  
In fact
it is known that for infinitely long ranged
interactions of a Kac form, the critical fluctuations
are not important and the mean-field theory becomes
exact. Apparently something very similar happens in the
case of Stockmayer fluids 
($DHS$ plus dispersion interaction).
As the net interaction between the particles becomes
more and more 
dominated by the long ranged dipole-dipole force, the critical
fluctuations become less and less important. 
For $0.3<\lambda<0.5$, this should also lead to a  
mean-field-like scaling in the neighborhood
of the critical point, 
with Ising-like behavior appearing
only very close to the critical point.  In fact
the width of the Ising-like critical region 
should be a very strongly
decreasing  function of $\lambda$.  It would be 
nice to see how these predictions 
compare with the Monte Carlo
simulations.

We have discussed two models of strongly correlated fluids:
the restricted primitive model of electrolyte and the
dipolar hard sphere model of polar fluid.
In both cases the bare interactions between particles are
long ranged, and the excess free energy
is dominated by the electrostatic correlations,
with the mean-field contribution being identically 
zero. These
are but two examples of an infinite number of
strongly correlated fluids.  
In spite of their ubiquity, the
critical behavior or even the existence of phase transition
in this important class of systems remains  
uncertain~\cite{Cam00,Tlu00,Pin00}.

This work was supported in part by  
Conselho Nacional de
Desenvolvimento Cient{\'\i}fico e Tecnol{\'o}gico (CNPq).

\end{document}